\documentclass[prl,aps,twocolumn,showpacs]{revtex4}
\newlength{\figwidth}
\usepackage{graphicx}
\begin{document}

\title{Spectral statistics of a quantum interval-exchange map}
\author{E. Bogomolny}
\author{C. Schmit}
\affiliation {\textit{Laboratoire de Physique Th\'eorique et Model\`es
Statistiques*\\ Universit\'e de Paris-Sud, B\^atiment 100, 91405 Orsay Cedex,
France}}

\date{\today}

\begin{abstract}
Curious spectral properties of an ensemble of random unitary matrices
appearing in the quantization of a map
 $p\to p+\alpha$, $q\to q+f(p+\alpha)$ in \cite{Marklof} are investigated. 
When $\alpha=m/n$ with integer co-prime $m,n$ and matrix dimension $N\to \infty$ is such that
$mN\equiv \pm 1$ mod $n$, local spectral statistics of this ensemble tends 
to the semi-Poisson distribution \cite{Gerland} with arbitrary
integer or half-integer level repulsion at small distances:
$R_2(s)\to s^{\beta}$ when $s\to 0$ and $\beta=n-1$ or
$n/2-1$ depending on time-reversal symmetry of the map.


\end{abstract}
\maketitle
Random matrices appear naturally in many different physical and mathematical
problems ranging from nuclear physics to number
theory (see e.g. \cite{Mehta}). Though many different ensembles of random matrices were
considered, it is well accepted \cite{Mehta} that the level repulsion (i.e. the
small-$s$ behaviour of the two-point correlation function
$R_2(s)\stackrel{s\to 0}{\longrightarrow} s^{\beta}$) is determined only from
symmetry arguments: $\beta=1,2,4$ for, respectively, real symmetric, complex
hermitian, and self-dual quaternion matrices (plus, of course, $\beta=0$
for diagonal matrices with independent entries). Dyson's arguments leading to
these values are so robust that only rarely one considers seriously the
possibility that certain models may have different values of $\beta$ (see e.g.
\cite{Robnik} ).

The purpose of this Letter is to demonstrate that a quantization of an
interval-exchange map leads to a random unitary matrix ensemble which, 
under certain conditions, reveals  level repulsion with arbitrary integer and
half-integer exponent $\beta$. To the authors knowledge it is the first
example which shows naturally level repulsion different from standard values
(but see \cite{Yan}).  

According to \cite{Marklof} a suitable quantization of the map
\begin{equation}
\Phi_{\alpha}: \left ( \begin{array}{c} p\\q\end{array}\right )\longmapsto
 \left ( \begin{array}{c} p+\alpha\\q+f(p+\alpha)\end{array}\right )\;
 \text{ mod }1
\label{map}
\end{equation}
(where $\alpha$ is a constant and $f(q)$ is a certain function with
period $1$) leads in the momentum representation to the following 
$N\times N$ unitary matrix ($k,p=0,1,\ldots,N-1$)
\begin{equation}
M_{kp}=e^{i\Phi_k}\frac{1-e^{2\pi i \alpha N}}{N(1-e^{2\pi i (k-p+\alpha
N)/N})}\ ,
\label{main}
\end{equation}
where $\Phi_k=-2\pi N F(k/N)$ and $F'(p)=f(p)$. 
Its eigenvalues have the form $\Lambda(j)=e^{ i\varphi(j)}$ with
real $\varphi(j)$ called eigenphases and we are interested in
their distribution.

To define an ensemble of random matrices it is convenient to consider
$\Phi_k$ in (\ref{main}) not as derived from a function $F(k)$ but as
independent random variables \cite{Gauss}. More precisely, the following
two cases are
considered. In the first one (called non-symmetric matrices) all  $\Phi_k$ with
$k=0,\ldots N-1$ are independent random variables distributed
uniformly between $0$ and $2\pi$. In the second case (called symmetric
matrices) only a half of $\Phi_k$ is independent. The others are obtained from
symmetry relations $\Phi_{N-k}=\Phi_k$.
These two cases correspond  to classical maps without and with 
time-reversal invariance.

When $\alpha$ is a `good' irrational number (e.g. $\alpha=\sqrt{5}/2$) 
numerical calculations show that  spectral statistics of symmetric 
(resp. non-symmetric) matrices  is well described for large $N$ by
the standard Gaussian orthogonal (resp. unitary) ensemble of random
matrices as it follows from usual symmetry considerations.

For rational values of $\alpha=m/n$ with
integer co-prime $m$ and $n$ only a finite number of different momenta  appear 
under the iterations and classical map (\ref{map}) can be identified with 
an interval-exchange map. 
  
\begin{figure}[ht] 
\begin{center} 
\includegraphics[height=\figwidth, angle=270]{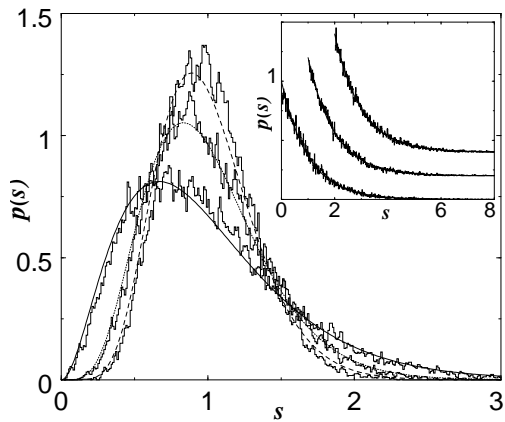} 
\end{center} 
\caption{The nearest-neighbor distribution for 
non-symmetric matrices with $\alpha=1/3, 1/6, 1/9$ and resp. 
$N=202, 205, 206$. 
Solid, dotted, and dashed lines are prediction (\ref{pbeta}) with resp. 
$\beta=2, 5, 8$. Insert: the same but for $M^n$ where $n$ is the denominator
of $\alpha$. For clarity pictures are shifted along the diagonal. 
Solid lines: the Poisson prediction $p(s)=e^{-s}$.
\label{fig1}} 
\end{figure}

\begin{figure}[ht] 
\begin{center} 
\includegraphics[height=\figwidth, angle=270]{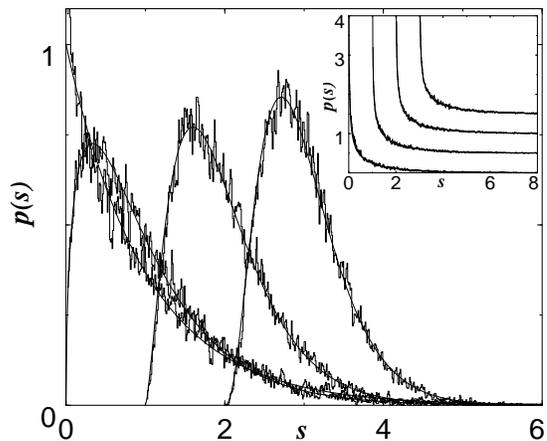} \end{center} 
\caption{The nearest-neighbor distribution for symmetric matrices with 
$\alpha=1/2, 1/3, 1/5, 1/7$ and resp. $N=201, 202, 201, 202$. For clarity 
pictures for $\alpha=1/5$ and $1/7$ are shifted horizontally by 1 and 2 units. 
Solid lines are prediction (\ref{pbeta}) with resp. $\beta=0,1/2,3/2,5/2$. 
Insert: the same but for $M^n$. Solid lines: the super-Poisson distribution 
(\ref{super}).
\label{fig2}} 
\end{figure}

Our main result in this case is the following. When $\alpha=m/n$ and 
matrix  dimension $N\to \infty$ is such that
\begin{equation}
mN\equiv \pm 1 \;\text{ mod }n
\label{condition}
\end{equation}
local spectral statistics of matrix ensemble (\ref{main})  is described by 
the semi-Poisson
statistics characterized by a parameter $\beta$ related with $\alpha$ as follows
\begin{equation}
\beta=\left \{ \begin{array}{cl}n-1&\text{ for non-symmetric matrices}\\
n/2-1&\text{ for symmetric matrices}\end{array}\right . \ .
\label{beta}
\end{equation}
The general semi-Poisson statistics was investigated in \cite{Bogomolny} and 
\cite{Gerland} as the simplest model of intermediate
statistics. It is defined in such a way that the probability of 
having $N$ ordered levels $E_1\leq E_2\leq \ldots \leq E_N$ is
\begin{equation}
p_{\beta}(E_1,\ldots,E_N)\sim \prod_{j=1}^{N-1}|E_{j+1}-E_{j}|^{\beta}\ .
\end{equation}
For this model all correlation functions are known when $N\to \infty$  
\cite{Gerland}. In particular, the nearest-neighbor 
distribution is
\begin{equation}
p_{\beta}(s)=A_{\beta}s^{\beta}e^{-(\beta+1)s}\ , \;
A_{\beta}=\frac{(\beta+1)^{\beta+1}}{\Gamma(\beta+1)}\ ,
\label{pbeta}
\end{equation}
the two-point correlation form factor takes the form
\begin{equation}
K_{\beta}(\tau)=1+2\text{ Re }\frac{1}{(1+2\pi i \tau/(\beta+1))^{\beta+1}-1}
\ ,
\label{kt}
\end{equation}
and the level compressibility which determines the asym\-ptotic
behaviour of the number variance is $\chi\equiv K(0)=1/(\beta+1)$. 
The two-point correlation function is expressed through the
Mittag-Leffler function \cite{Bateman}. For integer $\beta=n-1$ it is a finite
 sum
\begin{equation}
R_2^{(n)}(s)=e^{-n s}\sum_{k=0}^{n-1}\exp\left [n se^{2\pi
i k/n}+2\pi i \frac{k}{n}\right ]\ .
\label{r2beta}
\end{equation}
In Figs.~\ref{fig1} and \ref{fig2} a few examples of nearest-neighbor 
distributions
for quantum maps with different $\alpha$ and symmetries are 
presented. For each matrix 100 different realizations of
random phases were considered and in all figures the spectrum is 
unfolded to unit mean density. Prediction (\ref{pbeta}) with (\ref{beta})
agrees  very well for all $\alpha$ with numerical calculations.

We sketch main steps leading to the above result. Details will be given 
elsewhere \cite{inprogress}. First, for all $\alpha$ and $N$ eigenphases of 
matrix 
(\ref{main}) are such that for all nearby pairs of eigenphases $\varphi_1$ and 
$\varphi_2$ there is one and only one eigenphase $\varphi$ such that 
$\varphi+2\pi \alpha$ falls in-between $\varphi_1$ and $\varphi_2$ and there 
is one and only one $\varphi'$ such that the same is true for 
$\varphi'-2\pi \alpha$ 
\begin{equation}   
\varphi_1\leq \varphi+2\pi \alpha \leq \varphi_2\ ,\ 
\varphi_1\leq \varphi'-2\pi \alpha \leq \varphi_2\ . \label{rank} 
\end{equation} 
To check it let us consider instead of matrix $M_{kp}$ (\ref{main}) a new 
matrix 
\begin{equation} 
N_{kp}=M_{kp}e^{2\pi i (k-p+\alpha N)/N}\ . 
\label{nkp} 
\end{equation} 
If $u_k(j)$ and $\Lambda(j)$ for 
$j=1,\ldots ,N$ are eigenfunctions and eigenvalues of matrix $M_{kp}$ 
($\Lambda(j)u_k(j)=\sum_pM_{kp}u_p(j)$) then  $\Psi_k(j)=e^{2\pi i k/N}u_k(j)$ 
and $\Lambda'(j)=e^{2\pi i\alpha}\Lambda(j)$ are eigenfunctions and 
eigenvalues 
of matrix $N_{kp}$. On the other hand, due to special form of matrix $M_{kp}$, 
matrix $N_{kp}$ (\ref{nkp}) is a rank-one perturbation of matrix $M_{kp}$ 
\begin{equation} N_{kp}=M_{kp}-\frac{1-e^{2\pi i \alpha N}}{N}e^{i\Phi_k}\ . 
\end{equation} It is known (see e.g. \cite{Albeverio}) (and can be  checked in 
this case) that for rank-one perturbations new eigenvalues are in-between the 
unperturbed ones. As  both are known, the first inequality in (\ref{rank}) 
follows. The second can be proved by similar arguments. 

These inequalities manifest the existence of long
distance correlations between eigenphases of matrix (\ref{main}).
When condition (\ref{condition}) is fulfill, these long correlations
induces a very particular short ordering.
Namely, consider, instead of matrix $M$ defined by (\ref{main}) its 
$n^{\text{\scriptsize th}}$  power, $M^n$, where $n$ is the denominator of $\alpha$. 
Of course, eigenvalues of this matrix are  just the  $n^{\text{\scriptsize th}}$ power of eigenvalues 
of matrix $M$ but, as eigenphases are defined only modulo $2\pi$, new 
eigenphases are equal to $n\varphi(j)$ mod $2\pi$. Therefore, even if between
 two eigenphases 
$\varphi_1$ and $\varphi_2$ there were no eigenphases of matrix $M$, 
between $n\varphi_1$ and $n\varphi_2$ in $M^n$ there will be, in general, a few new 
eigenphases coming from different numbers of rotations around the circle. 
Nevertheless, condition (\ref{condition}) is sufficient to insure that in 
matrix $M^n$ in-between $n\varphi_1$ and $n\varphi_2$ there exist exactly $n-1$
new eigenphases. 

To prove it we put all eigenphases of matrix $M$ on the unit circle and
divide it in $n$ consecutive sectors of angle $2\pi \alpha=2\pi m/n$. Let $n_k$ be
the number of eigenphases inside the $k^{\text{\scriptsize th}}$ sector. 
Denote by $x_k$ and 
$y_k$ the (positive) distances between angle $2\pi m (k-1)/n$ (i.e. the beginning 
of the $k^{\text{\scriptsize th}}$ sector) and the two eigenphases closest to this angle 
in such a way that $x_k$ corresponds to an eigenphase in the 
$k^{\text{\scriptsize th}}$
sector and $y_k$ to one in $(k-1)^{\text{\scriptsize th}}$ sector. Inequalities 
(\ref{rank}) lead to the following inequality for all $k$ \cite{ignore} 
\begin{equation}
(y_{k+1}-y_k)(x_{k+1}-x_k)<0_ .
\label{boundary}
\end{equation}  
As in the $(k+1)^{\text{\scriptsize th}}$ sector it must exist eigenphases 
which after the shift by 
$-2\pi \alpha$  come in-between $n_k$ eigenphases of the 
$k^{\text{\scriptsize th}}$ 
sector, it follows that the number of eigenphases in the 
$(k+1)^{\text{\scriptsize th}}$ 
sector is at least 
$n_k-1$. The only possibility to get more eigenphases in the 
$(k+1)^{\text{\scriptsize th}}$ sector is connected with the positions of 
the last and the first eigenphases in the sector.
Straightforward calculations plus (\ref{boundary}) give 
recursive relations
\begin{equation}
n_{k+1}=n_k-1+\Theta(x_{k+2}-x_{k+1})+\Theta(x_{k}-x_{k+1})\ .
\label{recursive}
\end{equation}  
Here $\Theta(x)=1$ if $x>0$ and $\Theta(x)=0$ if $x<0$. 

Now choose the beginning of the first sector at the position of an eigenphase
(i.e. $y_1=0$). From (\ref{boundary}) it follows that $x_2<x_1$ and 
$x_n<x_1$. Direct applications of (\ref{recursive})
show that for $j=2,\ldots,n$ (with $x_{n+1}=x_1$)
\begin{equation}
n_j=n_1+\Theta(x_{j+1}-x_j)\ . 
\label{nj}
\end{equation}
As $\sum_{k=1}^n n_k=mN$ 
(because it corresponds to $m$ full turns around the unit circle) one obtains
\begin{equation}
mN=n n_1+1+\sum_{j=2}^{n-1}\Theta(x_{j+1}-x_j)\ .
\end{equation}
If $mN\equiv 1$ mod $n$, all the $\Theta$-functions have to be equal to 0
which leads to the following chain of inequalities
\begin{equation}
0<x_n<x_{n-1}<\ldots<x_2<x_1\ .
\label{order}
\end{equation}
But $x_1$ is just the distance between a pair of two nearby eigenphases of 
matrix (\ref{main}) and the true eigenphase corresponding to $x_k$ is 
$2\pi m(k-1)/n+x_k$. Distances of all other eigenphases from
boundaries of sectors are bigger than $x_1$.  When matrix $M^n$ is considered, 
multiples of $2\pi$ have to be ignored and one concludes that in-between two
eigenphases of matrix $M^n$ coming from two nearby eigenphases of matrix $M$
there are exactly $n-1$ other eigenphases ordered as in (\ref{order}).  
When  $mN\equiv -1$ mod $n$ all the $\Theta$-functions have to be equal to 1
which leads to the same conclusion (but $n-1$ eigenphases appear in the 
inverted order). In other words,  when eigenphases of matrix $M^n$ are known
and (\ref{condition}) is satisfied,  correct ordering of eigenphases of 
matrix $M$ 
corresponds simply to considering each $n^{\text{\scriptsize th}}$ eigenphase of $M^n$ (and dividing them by $n$). 

But the $n^{\text{\scriptsize th}}$ power of the classical map (\ref{map}) with 
rational $\alpha=m/n$ is classically integrable (as the momentum is conserved)
\begin{equation}
\Phi_{\alpha}^n: \left ( \begin{array}{c} p\\q\end{array}\right )
\longmapsto
 \left ( \begin{array}{c} p\\q+\sum_{j=1}^n f(p+j\alpha)\end{array}\right )\;
 \text{ mod }1\ .
\label{mapq}
\end{equation}
Quantization of this map as in \cite{Marklof} leads to a quantum map whose 
eigenphases are $\varphi_0(k)=-2\pi N \sum_{j=1}^n F(k/N+j\alpha)$. If in the 
semiclassical limit $N\to \infty$ classical behaviour 
dominates, it is natural that local spectral statistics of $M^n$ (for all $N$) 
will be Poissonian as it is conjectured  for all generic integrable models 
\cite{Tabor}. It seems that it is the case for non-symmetric matrices (with
an arbitrary function $f(k)$) because  all $\varphi_0(k)$  are
different and small corrections from omitted terms are negligible. But for
symmetric matrices $F(-k)=F(k)$, consequently, $\varphi_0(N-k)=\varphi_0(k)$
and these corrections are important as unperturbed eigenphases form degenerate
 pairs. 
Nevertheless, one can argue that each second eigenphase of $M^n$ (i.e.
either odd or even in the ordered sequence) still form a Poisson sequence
but degenerate levels are split. Let $p(s)$ be the distribution of
nearest-neighbor odd-even eigenphases with mean density 1. As
odd or even eigenphases of matrix $M^n$ are assumed to obey the Poisson
statistics (with density equals one half of the total density) one
gets a convolution equation
\begin{equation}
\frac{1}{2}e^{-s/2}=\int_0^s p(l) p(s-l)dl
\end{equation}
whose solution is
\begin{equation}
p(s)=\frac{1}{\sqrt{2\pi s}}e^{-s/2}\ .
\label{super}
\end{equation}
Higher correlation functions are obtained in the same manner 
and one concludes that the odd-even distribution  is
exactly the semi-Poisson distribution with $\beta=-1/2$ (cf. (\ref{pbeta})). For brevity we refer to it as to the super-Poisson distribution because instead
of level repulsion it shows level attraction 
($p(s)\stackrel{s\to 0}{\longrightarrow}\infty$) \cite{superPoisson}.

The above arguments demonstrate that for large $N$ eigenphases of matrix 
$M^n$ where $n$ is the denominator of $\alpha$  have universal distribution
(independent of $\alpha$), namely, the Poisson distribution for
non-symmetric matrices and the super-Poisson distribution (i.e. the semi-Poisson distribution with $\beta=-1/2$) for symmetric ones 
(see inserts in Figs.~\ref{fig1} and \ref{fig2}). 

When $N$ obeys (\ref{condition}) eigenphases of $M$ are obtained simply by
jumping on each $n^{\text{\scriptsize th}}$ eigenphase of matrix $M^n$ which
 leads
directly to (\ref{beta}). In \cite{Seligman} it was noted that for integer
$\beta$ the semi-Poisson distribution can be obtained from  a Poisson sequence
by considering each $(\beta+1)$ level. It is interesting that  this completely 
artificial  mechanism appears naturally in the quantization of the 
 map (\ref{main}) (for symmetric matrices one jumps
over not a Poisson sequence but a super-Poisson one). 

\begin{figure}[ht]
\begin{center}
\includegraphics[height=\figwidth, angle=270]{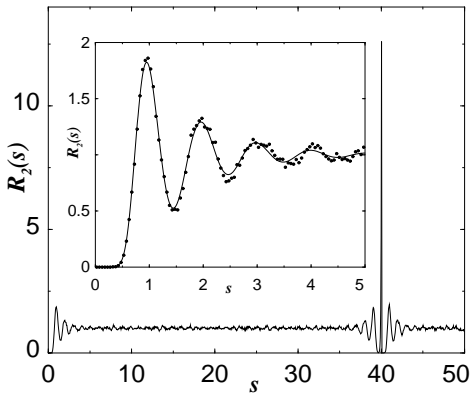} 
\end{center}
\caption{The two point correlation function for non-symmetric matrices with
$\alpha=1/20$ and $N=801$. Insert: small-$s$ behaviour of this quantity.
Solid line:  Eq.~(\ref{r2beta}) with $n=20$.
\label{fig5}}
\end{figure}

The described result applies only to local statistics in the universal limit
of unit mean density and $N\to \infty$. For long-range statistics
(on distances of the order of $N$ in this scale) the above-mentioned long-range correlations
between eigenphases of matrix $M$ become apparent. In Fig.~\ref{fig5} 
the two-point
correlation function for non-symmetric matrices with $\alpha=1/20$ and
$N=801$ is presented. In the insert the correlation function till $s=5$ is
given together with the prediction (\ref{r2beta}) with $n=20$. The very
strong level repulsion ($\sim s^{19}$) and prominent oscillations described
by this formula agree very well with the numerical calculations. Nevertheless,
for $s$ close to $\alpha N$ one observes a big peak which is a clear signal of
 long-range correlations. At higher $s$ new peaks (with smaller amplitudes) appear at multiples of this quantity. 

We also investigate numerically eigenfunctions $u_k(j)$ of matrix $M$. As
expected  for models with intermediate statistics they show fractal
properties which manifest e.g. in a non-trivial power increase of 
participation ratios
\begin{equation}
R_q=\langle \sum_k|u_k(j)|^{2q}\rangle_j 
\stackrel{N\to \infty}{\longrightarrow} N^{D_q(q-1)}
\end{equation}
where $\sum_k|u_k(j)|^{2}=1$ and $\langle \ldots \rangle_j$ denotes an
averaging over the spectrum.  For example, for non-symmetric matrices with 
$\alpha=1/2$,
$D_2\approx D_3\approx D_4\approx .4$. 

In conclusion, we investigate unusual spectral statistics of random unitary
matrices coming from a quantization of map (\ref{main}). 
The most interesting case corresponds to $\alpha=m/n$  (when the map is an 
interval-exchange map) and $mN\equiv \pm 1$ mod $n$ \cite{other}. When 
$N\to \infty$ local spectral statistics of these matrices tends to the
semi-Poisson statistics with parameter $\beta$ which can take
arbitrary integer and half-integer values depending on the denominator of 
$\alpha$ and the time-reversal invariance of the map. The level compressibility 
is non-zero \cite{chi}. Eigenphases of $M^n$ for all $N$ have the Poisson
distribution for non-symmetric matrices and the super-Poisson distribution
for symmetric ones. When $mN\equiv \pm 1$ mod $n$ eigenphases of $M$
are obtained by jumping over each $n^{\text{\scriptsize th}}$ eigenphase of 
$M^n$ thus 
explicitly realizing the  \textit{ad hoc} mechanism proposed  
in \cite{Seligman}.
Eigenfunctions have fractal properties in momentum representation.

The model considered seems to be  the first example of a new class of
random matrix ensembles and its detailed analysis may lead to a better
understanding of spectral statistics of pseudo-integrable plane polygonal
billiards whose classical mechanics is described by interval-exchange maps.

\begin{acknowledgments} The authors are greatly indebted to J. Marklof for
discussing the paper \cite{Marklof} prior the publication and for numerous
useful discussions and comments. One of the author (E. B) is thankful to 
O. Giraud for useful discussions and to Y. Fyodorov for pointing out 
Ref.~\cite{Yan}.
\end{acknowledgments}


\begin{thebibliography}{99}
\bibitem{Gauss} In \cite{Marklof} the case $\Phi(k)=k^2$ was considered but
the answer depends strongly on 
arithmetic nature of $N$ due to special properties of the Gauss sums.
\bibitem{ignore} For simplicity, degenerate cases like $x_{k+1}=x_{k}$ are 
ignored.
\bibitem{superPoisson} Let $x_1\leq x_2\leq \ldots \leq x_N$ form a Poisson
 sequence. Put in-between
$x_k$ and $x_{k+1}$ a new point $y_k=x_k+(x_{k+1}-x_k)(1-\cos \Theta_k)/2$
where $\Theta_k$ are independent random variables with uniform distribution
between 0 and $2\pi$. Then $x_k$ and $y_k$ together form the super-Poisson
sequence \cite{inprogress}.
\bibitem{other} Other cases are less drastic and will be considered elsewhere 
\cite{inprogress}.
\end{thebibliography}
\end{document}